\documentclass[superscriptaddress,nofootinbib,twocolumn,longbibliography]{revtex4-1}

\usepackage[bottom]{footmisc}
\usepackage[dvipsnames]{xcolor}
\usepackage{lmodern}
\usepackage{amsmath}
\usepackage{amssymb}
\usepackage{fontenc}
\usepackage{enumerate}
\usepackage{mathrsfs}
\usepackage{float}
\usepackage{graphicx}
\usepackage{subcaption}
\usepackage{soul}

\def\beq{\begin{equation}}
\def\eeq{\end{equation}}

\newcommand{\be}{\begin{equation}}
\newcommand{\ee}{\end{equation}}
\newcommand{\ben}{\begin{enumerate}}
\newcommand{\een}{\end{enumerate}}

\def\e{\epsilon}

\usepackage{hyperref}
\hypersetup{
    colorlinks=true,
    linkcolor=RoyalPurple,
    filecolor=RoyalPurple, 
    citecolor=RoyalPurple,
    urlcolor=RoyalPurple,
    }
    
\begin{document}

\title{On the lapse contour\\
in the gravitational path integral}

\author{Batoul Banihashemi}
\email{bbanihas@ucsc.edu}
\affiliation{Physics Department, UC Santa Cruz, CA 95064, USA}

\author{Ted Jacobson}
\email{jacobson@umd.edu}
\affiliation{Maryland Center for Fundamental Physics,\\  University of Maryland, College Park, MD 20742, USA}

\begin{abstract}
The gravitational path integral is usually implemented with a covariant action by analogy with other gauge field theories, but the gravitational case is different in important ways. A key difference is that the integrand has an essential singularity, which occurs at zero lapse where the spacetime metric degenerates.  The lapse integration contour required to impose the local time reparametrization constraints must run from $-\infty$ to $+\infty$, yet must not pass through zero. This raises the question: 
for an application---such as a partition function---where the constraints should be imposed, 
what is the correct integration contour, and why? We study that question by starting with the reduced phase space path integral, which involves no essential singularity. We observe that if the momenta are to be integrated before the lapse, to obtain a configuration space path integral, the lapse contour should pass below the origin in the complex lapse plane. This contour is also consistent with the requirement that quantum field fluctuation amplitudes have the usual short distance vacuum form, and with obtaining the Bekenstein-Hawking horizon entropy from a Lorentzian path integral.
\end{abstract}

\maketitle

\section{Introduction}

The gravitational path integral \cite{Misner:1957wq} seems to offer some 
insight into non-perturbative aspects of quantum gravity,
despite its UV incompleteness. Beginning with the seminal
work of Gibbons and Hawking \cite{Gibbons:1976ue} on black hole and de Sitter 
horizon entropies, it has been applied in numerous ways,
to various species of black holes and boundary conditions, 
higher derivative corrections to the gravity theory, 
phase transitions, quantum cosmology, and holographic entanglement
entropy, to name a few. In all of these applications, it is 
a saddle point approximation to the covariant path integral that 
plays the central role, most often at a Euclidean saddle.

Unlike in the case of non-gravitational quantum field theories,
the derivation of a covariant gravitational path integral starting from 
a (reduced) phase space path integral (which takes canonical
quantization as its starting point) presents complications and
questions that have yet to be fully resolved, stemming from the 
nature of the diffeomorphism gauge invariance, and the 
fact that the nondegenerate Riemannian spatial
metrics do not
form a linear space so canonical quantization is not strictly 
applicable. From a practical 
viewpoint, without attempting to resolve all such questions 
at this stage, it is imperative at least to identify  provisionally suitable integration variables, 
their contour of integration, and whether that
contour can be deformed so as to access the saddles that are 
thought to yield physically valid approximations.

At the classical level, the lapse functions as a Lagrange multiplier in the
action, whose variational equation imposes the Hamiltonian constraint
(which generates spatial hypersurface deformations). 
It is related to the $tt$ component of the
inverse metric via $N =  1/\sqrt{-g^{tt}}$. 
If the path integral is employed for state counting 
via a trace over the physical Hilbert space---which is the motivation behind our 
analysis in this paper---the constraints should be imposed on the paths.\footnote{The 
gravitational path integral may instead be used for other purposes, 
such as computing a ``Feynman propagator", or a wavefunction that satisfies 
the Wheeler-DeWitt equation (for example the no-boundary wavefunction). 
In such applications, imposition of the constraints on the paths may not be required, and a different lapse contour may be called for \cite{Teitelboim:1983fh, Teitelboim:1981ua, Feldbrugge:2017kzv}.}
And in order for the lapse integral to impose the Hamiltonian constraint, 
its range of integration must include
the whole real line.
In a covariant gravitational path integral over
the metric components, one must therefore 
in addition sum over the two signs
of the volume density $\sqrt{-g}$ in order to include
both signs of $N$. (Other variable choices are
mentioned in the Discussion section.)
The integration range thus includes 
degenerate spacetime metrics where the action of general relativity is ill defined since $N=0$.

To untangle this knot, we take
the starting point to be the presumably more fundamental reduced
gravitational phase space path integral, 
as formulated by Faddeev and Popov \cite{Faddeev:1973zb},
and we focus in particular on the character and contour of the
integral of the Lagrange multiplier that
is identified with the lapse.
By careful consideration of the order of integration,
we find that the price of reformulating it as a covariant integral is
that the lapse itself, rather than the $g^{tt}$ metric component, must be
taken as the field variable, and its integration contour must navigate 
around the origin in a particular direction in the complex plane.
This complex lapse contour is thus derived from first principles.
Moreover,
 it  satisfies the fluctuation convergence criterion first 
 formulated by Halliwell and Hartle \cite{Halliwell:1989dy} and more recently discussed by 
others \cite{Louko:1995jw,Kontsevich:2021dmb,Witten:2021nzp}, 
and in a simplicial minisuperspace Lorentzian computation of horizon 
entropy it can be deformed to pass through  the Euclidean saddle that yields the Bekenstein-Hawking entropy~\cite{Dittrich:2024awu}. 
It also has implications for 
path integral definitions of a 
Wheeler-DeWitt wave function for quantum cosmology.

In a related study in a minisuperspace context \cite{Marolf:1996gb}, Marolf 
considered matrix elements of 
a Fourier representation of a 
delta function of the Hamiltonian constraint,
expressed as a phase space path integral,
and carried out the momentum integrals first, 
interpreting them as defining
a distribution. 
In contrast, we have determined the 
deformation of the
original lapse contour, which was along the real line, 
so as to allow the interchange of the 
order of lapse and momentum integration.

\section{Path integral and lapse contour}

We begin with the phase space path integral, since that is 
constructed to be equivalent to canonical quantization, which 
we take to be the definition of the quantum theory.\footnote{In general
relativity, even aside from the issue of UV incompleteness, 
standard canonical quantization is not admissible. The
canonical commutation relations imply that the 
momentum is the generator of metric translations,
so the metric components could be translated to 
values that violate the Euclidean
signature restriction on the spatial metric. Suitable variants 
of canonical quantization such as affine quantization~\cite{Klauder:2022byn,Klauder:2023woi} could 
presumably take care of this complication, but for our present purposes
we shall ignore it, since it hopefully has no impact on the 
issue we are addressing here. Moreover, it may well
be that it cannot be correctly addressed without a UV completion
of the theory.}
The phase space path integral
for the propagator
of a quantum system with Hamiltonian $H$, i.e., the matrix elements of the (retarded) time evolution
operator, takes the form 
\be \label{canonical-PI}
\int \underset{t,i}{\prod}dp_i(t) dq^i(t)\exp\left(i\int dt\, (p_i \dot q^i - H)\right)
\ee
with suitable boundary conditions (and $\hbar =1$). 
The integration measure is over the paths 
of canonical coordinates in the physical phase space 
of true degrees of freedom. 
For a constrained Hamiltonian system,
such as a gauge theory or general relativity, canonical quantization
is defined on the reduced phase space, in which the constraints  
are satisfied and the gauge freedom has been fixed. 
Faddeev and Popov~\cite{Faddeev:1973zb}
showed  that the path integral on the reduced phase space can 
be written as a path integral on the {\it un}reduced 
phase space, by including at each time step in the integration measure 
delta functions imposing the constraints ${\cal C}^a$ and
gauge fixing functions $\chi_a$, together with the determinant
of their Poisson bracket,  viz., 
$\underset{a}{\prod}\delta({\cal C}^a)\delta(\chi_a){\rm det}\{\chi_b,{\cal C}^c\}$.
(In a field theory the constraint and gauge fixing indices denote
the point in space as well as any other multiplicity of the 
constraints.)

In general relativity,
when lifting to the unreduced phase space, the 
$H$ term in \eqref{canonical-PI} 
takes the form of  
additional $p_a \dot q^a$
terms (up to a possible boundary term that is not
relevant to the present discussion), so the bulk Hamiltonian
disappears.  However,
if the Hamiltonian constraint delta functions 
$\delta({\cal H})$
at each time step
are written
as Fourier integrals at each point in space,
\be\label{deltaH}
\delta({\cal H}) = \int \underset{x}{{\prod}}\frac{dN}{2\pi} \exp\left(-i\int N{\cal H}\right) \, ,
\ee
a Hamiltonian term in 
the unreduced phase space 
action 
appears in the guise of
the integral
$\int N {\cal H}$.
The Lagrange multiplier field $N$ thus plays the
role of the lapse
function of the classical 
ADM decomposition of the 
spacetime metric \cite{Arnowitt:1962hi}.
A suitable lapse contour for 
imposing the constraint
runs from $-\infty$ to $\infty$ along
the real line, provided the lapse integral
is carried out {\it before} those of the
phase space coordinates. If, however, 
we wish to first integrate the momenta so as to arrive at a covariant path integral, the lapse contour must be chosen
more judiciously. Our aim here is to determine how that should be done.

The Hamiltonian constraint in vacuum general relativity 
(at each spatial point) in $3+1$ spacetime dimensions
and in units with $16\pi G =c=1$
is 
\begin{align}\label{calH}
    {\cal H} &=  q^{-\frac12}(p^{ij}p_{ij}-\tfrac12 p^2) - q^{\frac12}R\,,
\end{align}
where $q$ denotes the determinant of the spatial metric $q_{ij}$,
$p^{ij}$ is the conjugate momentum, $p$ is its trace, 
the indices on $p_{ij}$ are implicitly 
lowered by contraction with $q_{ij}$,
and $R$ is the Ricci scalar of $q_{ij}$. 
The shift vector components also enter as Lagrange multipliers accompanying the momentum (spatial diffeomorphism) constraint; however, we do not focus on the shift in the following since that constraint is linear in $p^{ij}$, so does not affect the convergence of the integral.

To pass to a configuration space path integral
for a typical quantum system one performs the Gaussian integral 
over the momenta that appear quadratically in the kinetic energy, which has the effect of replacing the momenta in the exponent by their expression in terms of the time derivatives of the configuration
coordinates (and if the kinetic term in the Hamiltonian involves
those coordinates also introduces a nontrivial measure factor).
The momentum Gaussian has an imaginary variance, so the integral is not
absolutely convergent, but it can be made so by adding a positive real part 
$\e$ to the variance, and after doing the momentum integral  taking the limit $\e\to0$.\footnote{The physical justification of this mathematical maneuver, 
not typically mentioned, relies on the fact that one is interested in using the path integral to propagate physical states whose momentum dependence goes to zero as momentum goes to infinity. The only 
mention of this point we are aware of is in
Steven Weinberg's book on quantum field theory \cite{Weinberg_1995}, 
in the context of the path integral for vacuum expectation values.
We would appreciate being informed of other discussions in the context of quantum mechanics.}

In the case of gravity, we face the awkward fact that the constraint 
\eqref{calH} has both positive and negative 
quadratic terms in the momenta, so if the integral over the 
tracefree part of  $p^{ij}$ were to converge, that over the trace part 
would diverge.\footnote{In terms of the tracefree part $\hat p^{ij}:=p^{ij} - \frac13 p q^{ij}$ and the trace $p$, 
the quadratic expression in \eqref{calH} takes the form 
$\hat p^{ij}\hat p_{ij} - \frac16 p^2$, so indeed the squared trace 
enters with a minus sign.}
This appears at first to be a serious obstruction to integrating out the momenta, however
the gauge fixing condition could rescue us. There are of course 
innumerable ways to fix the gauge, but if one of them could permit
the momentum integration, 
that might be sufficient to 
establish equivalence with a path integral governed by a covariant action along with any 
gauge fixing and corresponding ghosts.
Here we will be content to note that
if the trace $p$ is gauge-fixed
the minus sign becomes harmless
for the remaining momentum integrals.

As an example, suppose that we impose the gauge condition $\chi = p=0$, which 
is equivalent to requiring that the time foliation consists of hypersurfaces with zero mean extrinsic curvature, i.e., maximal surfaces. Such a gauge 
does not exist globally in a generic spacetime, but for our present purposes 
it seems sufficient that it could generically 
be imposed locally in spacetime.\footnote{More generally, it would suffice to gauge-fix $p$ 
to a function of the configuration variables.} 
The Faddeev-Popov determinant 
involves the Poisson bracket of $\chi$
with the constraints. The bracket with 
the momentum constraints is zero, since
spatial diffeomorphisms preserve the vanishing of a spatial scalar density. 
The bracket with the Hamiltonian constraint can be easily computed 
using the Einstein equations in Hamiltonian form~\cite{Wald:1984rg},
and is given 
on the constraint surface
in vacuum general relativity by 
\begin{equation}\label{pHbracket}
    \{p(x),{\cal H}(y)\} = 2q^{1/2}\Bigl(R - (D_q^y)^2\Bigr)\delta(x,y)
\end{equation}
 where $D_q^y$ is the spatial metric-compatible covariant derivative operator 
acting on tensors at $y$.
This bracket is independent of the momenta $p^{ij}$,
so the momentum integrals remain simple Gaussians in this example.\footnote{The Poisson bracket with $p\equiv q_{ij}p^{ij}$ generates conformal rescalings of the spatial metric and inverse rescalings of the conjugate momentum. Since the $q^{-\frac12}(p^{ij}p_{ij}-\tfrac12 p^2)$ term in \eqref{calH} is homogeneous in both $q_{ij}$
 and $p^{ij}$, its contribution to $\{p(x),{\cal H}(y)\}$ is proportional to itself. On the constraint surface this contribution can be replaced by the negative of the remaining terms in 
 ${\cal H}$, and this is why \eqref{pHbracket} contains no terms quadratic in the momentum. (For spatially flat, homogeneous minisuperspace cosmology with no cosmological constant or matter field potential energy, this bracket vanishes, so $p$ is gauge invariant and hence cannot be gauge-fixed; however, this is a peculiarity of that symmetry-reduced theory.)}

Assuming now that some gauge condition 
eliminating
the potential
$p$ integral divergence
has been implemented,
the remaining momentum
integrals are convergent if the lapse is replaced by $N - i\e$ in
\eqref{deltaH}, yielding 
\be\label{deltaHe}
\delta({\cal H}) = \lim_{\e\to0^+}\int \underset{x}{{\prod}}\frac{dN}{2\pi} \exp\left( - i\int (N -i\e){\cal H}\right)\, .
\ee
Our proposal is that the order of integration can then be interchanged,
so the momentum integrals are performed first, but only provided that the limit $\e\to0^+$ is postponed until after that
integral is evaluated.
The presence of $\e>0$ is equivalent to an 
infinitesimal negative imaginary displacement of 
the lapse contour 
below the real line 
in the complex lapse plane. Without this displacement, the momentum integrals
would result in an integrand with an essential singularity at $N=0$,
rendering the subsequent lapse integrals 
ill defined.
This reasoning shows that the lapse contour displacement required to evade the essential singularity is 
not {\it ad hoc}, but rather is dictated by the imposition
of the constraint ${\cal H} = 0$, which is required in order to quantize the
physical degrees of freedom.   

To verify that with this contour
displacement the 
lapse integral can indeed be interchanged with the
momentum integrals 
we consider here an
illustrative example. 
The gravitational case is 
analogous to the following integral
over an $n$-dimensional momentum vector $p_i$
that satisfies a constraint on its norm:
\begin{align}
&\int d^np\; e^{ip_ix^i} \delta(p^2 - m^2)\nonumber\\& =  \tfrac12 m^{n-2}\Omega_{n-2}\int_{-1}^1 du\,(1-u^2)^{(n-3)/2} \cos(m |x| u)\label{n}\\
&= m^{n-2}  2^{\frac{n}{2}-1}\pi^{\frac{n}{2}} J_{\frac{n}{2}-1}(m|x|)/{(m|x|)^ {\frac{n}{2}-1}}\,.\label{nJ}
\end{align}
Here $p^2 \equiv p_ip_i$, and $\Omega_{n-2}$ is 
the surface area of the unit $(n-2)$-sphere.\footnote{As written,
\eqref{n} makes sense only for $n\ge2$ (with $\Omega_0=2$).
The case $n=1$ requires a different treatment than \eqref{n}, but it 
is correctly given by \eqref{nJ} with $n=1$.} 
In the first equality we have used hyperspherical polar coordinates for $p_i$, used 
the delta function to evaluate the integral over the norm of $p_i$, and set $u=\cos\theta$, where $\theta$ is the angle between $p_i$ and $x^i$.
The second equality is obtained from Poisson's integral
for the Bessel function.\footnote{See (10.9.4) of the NIST Digital Library of Mathematical Functions, \url{https://dlmf.nist.gov/10.9.E4}}.

If the delta function is written as a Fourier 
integral over a ``lapse'', the integral to be evaluated takes the form
\beq\label{n2}
\lim_{\e\rightarrow0^+} \int d^np\, (dN/2\pi)\; e^{ip_ix^i}e^{ - i(N - i\e)(p^2 - m^2)},
\eeq
where the $N$ contour runs along the real line.
If we do the $N$ integral first 
we may then take the limit $\e\to0^+$
and use the known distributional 
properties of the resulting delta function to evaluate
the $p_i$ integrals, as we did in \eqref{n}. However, 
if we want to first perform the $p_i$ integrals, then 
in order for those 
integrals to be convergent
the limit must be postponed,
and we obtain 
\beq\label{n3}
\lim_{\e\rightarrow0^+}\int \frac{dN}{2\pi}\, \left(\frac{\pi}{i(N - i\e)}\right)^{n/2}\; e^{ix^2/4(N-i\e)} e^{i(N-i\e)m^2},
\eeq
where  $(i(N - i\e))^{-n/2}$ takes its
principal value.
Note that if at this stage 
$\e$ is set to zero, the integrand develops an essential singularity
on the integration contour at $N=0$, 
so the $N$ integral becomes ill defined. 
The $i\e$ terms in \eqref{n3} are equivalent to the 
prescription that the $N$ contour is displaced by $-i\e$ 
below the real axis.

To check the agreement of \eqref{n3} with \eqref{nJ}
we change the integration variable to 
$t:= i(N-i\e)m^2$, 
in terms of which \eqref{n3} becomes
\beq\label{1}
\lim_{\e\rightarrow0^+}
\frac{- i m^{n-2}\pi^{n/2}}{2\pi} \int dt\, t^{-n/2}\;  \exp(t-{z^2}/{4t}) ,
\eeq
where $z = m|x|$, and the $t$ contour runs from $- i\infty +\e$ to $i\infty +\e$.
Since the arc at infinity contributes nothing, we may 
fold the contour so that it runs just 
below
the negative real axis, around the origin, and returns just 
above the negative real axis.
The integral in \eqref{1} is thus equal to 
$2\pi i (z/2)^{1-n/2}$ times
Schl\"afli's integral
for the Bessel function 
$J_{\frac{n}2-1}(z)$,\footnote{See
(10.9.19) in the DLMF, \url{https://dlmf.nist.gov/10.9\#Px7}, 
and Fig.\ 5.9.1 for the contour,
\url{https://dlmf.nist.gov/5.9.F1}}
hence the quantity in \eqref{1} is equal to that in \eqref{nJ}. 

The lesson is that 
indeed the lapse integral can be 
postponed until after the $p_i$ integrals, but only if 
the lapse contour is displaced so as to navigate in the complex plane
{\it below} the essential singularity at $N=0$. 
 Although we achieved this by a uniform displacement by $-i\e$, the oscillatory momentum integrals are conditionally convergent for $N$ different from zero, so if desired 
the contour can be pushed
back onto the real line except for a small detour around the origin in the complex $N$ plane. 
This elementary example lends strong support to our proposal 
that the contour for the lapse in the gravitational 
configuration space path integral
should run 
below the origin in the complex plane, 
thus passing around the essential singularity that occurs in
that context.

\section{Discussion}

We have argued that there is really no choice about the lapse contour, if one wishes to both impose the gravitational Hamiltonian constraint
at the level of the underlying Hilbert space\footnote{Another sense of imposing the constraint could be to impose the Wheeler-DeWitt equation on the wave function of the Universe, for which other contours could be used~\cite{Halliwell:1989dy, Halliwell:1990tu}.}
{\it and} to integrate the gravitational momenta 
prior to the lapse, with the aim of arriving at a covariant path integral. The contour 
should run just below the real line
in the complex lapse plane, or on any 
path related to that by an allowed deformation.\footnote{Had we written the Fourier representation of the delta function \eqref{deltaH} with the opposite sign in the exponent, we would have concluded that the $N$ contour should run 
{\it above} the origin rather than below.
This is equivalent to simply using the variable $-N$ instead of $N$. The variable choice we made has the 
property that, when $N>0$, $N{\cal H}$ 
generates hypersurface deformations in the direction of larger $t$ coordinate.}

This contour 
has two additional 
virtues besides allowing the interchange of the integration order.
One is that it renders 
convergent the integrals over matter field fluctuations as well, since the energy density of those fluctuations enters the Hamiltonian constraint 
in the same way as do the gravitational momenta.\footnote{There remain of course 
UV divergences associated with these integrals, but they are absorbed in renormalization counterterms.} 
Thus it satisfies the Hamiltonian version of the fluctuation convergence criterion of Halliwell and Hartle~\cite{Halliwell:1989dy}. 
The second virtue of this contour
is that it yields what is believed to be the correct
result for horizon entropy, as seen recently in two different settings.
In Refs.~\cite{Colin-Ellerin:2020mva,Marolf:2022ybi} horizon entropy 
was computed using a ``Lorentzian'' path integral, with a complex contribution
to the action for a Lorentzian cone singularity. The sign of the imaginary
part of metric needed to obtain the action that yields the correct entropy
is the same as that 
required by the fluctuation convergence criterion \cite{Louko:1995jw}.
Moreover, in Ref.~\cite{Dittrich:2024awu}
in the context of a simplicial minisupersapce 
path integral computation of horizon entropy, 
it was found that this contour can be deformed to pass 
through the Euclidean saddle that 
yields the correct, exponentially enhanced Bekenstein-Hawking entropy, rather than the suppressed exponential that would be obtained from the 
contour that passes on the other side of the essential singularity~\cite{Dittrich:2024awu}. 
These two virtues are not unrelated: the horizon entropy can be seen as a 
quantum gravity regulated 
UV completion of the otherwise divergent entanglement entropy of gravitational and matter fields in their local short distance vacuum state \cite{Solodukhin:2011gn}, which is associated with the infinite 
redshift at the horizon where $N\rightarrow0$ in the 
saddle configuration.

It has previously been argued \cite{Feldbrugge:2017mbc} 
that the contour advocated here is 
problematic in the context of the path integral for a state in quantum cosmology. That  argument was based on the observation that 
the contour can be deformed to traverse saddles 
around which field fluctuations are governed by an inverted Gaussian action,
which is inconsistent with semiclassical spacetime. 
If correct, this would imply that the lapse contour in a covariant path integral for the state in quantum cosmology cannot impose the constraints.
However, the claim that the
contour can be so deformed is highly questionable, because its conclusion
invalidates its premise. That is, if fluctuations were truly
governed by an inverted Gaussian, the quantum fields would be 
in a state that is nothing like the usual local vacuum at short distances,
so there would be an enormous (formally infinite) back-reaction, completely distorting the effective action in the neighborhood 
of what was the background saddle, and quite plausibly invalidating the contour deformation analysis.\footnote{The issue of
the back-reaction was addressed in Ref.~\cite{Feldbrugge:2017mbc}, but the 
fact that the state of the fluctuations would not 
coincide with that of the local vacuum at
short distances
was not taken into account.
This fact means that the usual divergences in the effective action
could not be absorbed in local counterterms, so that standard energy-momentum
tensor renormalization methods would fail, and semiclassical analysis would
fail to be applicable.} 

Where does our observation 
concerning the lapse contour 
leave things with respect to the
existence of a covariant, Lorentzian
gravitational path integral? 
We note first that 
the fact that the lapse must contain an infinitesimal negative imaginary part, at least
near the origin, means that strictly speaking the metric signature in the covariant path integral will be complex, rather than Lorentzian. However, the limit can and should be taken in which the imaginary part goes to zero, and in that limit the ``Lorentzian'' moniker applies.
Another caveat is that, since the lapse is
the {\it square root} of a metric component,\footnote{This is strictly true only
if the shift vector vanishes. Otherwise we should say that the lapse is the reciprocal of the square root of (minus) the time-time component of the inverse metric.}
the integration variables in 
a covariant path integral should evidently not 
be just the metric components. 
One way to 
implement the lapse integral 
covariantly would be to sum over both signs 
of $\sqrt{-g}$. Another way, followed by 
Faddeev and Popov~\cite{Faddeev:1973zb}, is to use the inverse metric density
variable $\sqrt{-g}g^{\mu\nu}$, whose $tt$ component is 
identified with the reciprocal of the lapse, multiplied by the square root of
the determinant of the spatial metric.
Alternatively, one could 
use the
tetrad formalism, integrating over both signs of the tetrad.
Yet another issue is that the
Euclidean signature and nondegeneracy of the spatial metric
should be preserved.
The triad formalism would
build in the spatial Euclidean signature,
but canonical quantization would allow
degenerate metric configurations which should
presumably be excluded.
On the other hand, perhaps that 
issue cannot be meaningfully addressed 
without a UV completion of the effective field theory.

We have discussed our analysis in the framework of
the phase space path integral, as formulated by 
Faddeev and Popov \cite{Faddeev:1973zb}, since that hews close
to the Hilbert space structure of the canonically
quantized theory. However, that is a gauge-fixed
formalism, and is not suited to implementation
of the renormalization in a manner that can guarantee
the preservation of gauge invariance and unitarity.
This limitation is plausibly irrelevant for the 
saddle point approximation, but it may render 
that formalism inadequate as a foundation for the 
effective field theory of quantum gravity.
An alternative formalism, introduced by 
Fradkin and Vilkovisky \cite{Fradkin:1973wke} (see also \cite{Fradkin:1975sj}),
includes ghost fields and BRST symmetry in order to
work with covariant gauge fixing and a local action 
whose renormalization is under control. In that
formulation, among the variables of integration are the
lapse and shift, and the lapse contour runs over the entire
real line. The essential singularity at vanishing lapse must
exist also in that formulation, hence the lapse contour must
not in fact run though the origin. Since the canonical
momenta have effectively been integrated out to
arrive at the covariant path integral of \cite{Fradkin:1973wke},
we expect that our conclusion about the lapse
contour applies to that path integral as well.

\vspace*{3mm}

\acknowledgments
We are grateful to Bianca Dittrich, Don Marolf, 
and Jos\'{e} Padua Arg\"{u}elles for helpful discussions. 
The work of BB was supported by 
DOE grant DE-SC001010 and the Federico and Elvia Faggin Foundation.
The work of TJ was supported in part by 
NSF grant  PHY-2309634.

\bibliography{Lapse-ref}

\end{document}